\documentclass[aps,prd,twocolumn,groupedaddress, showpacs]{revtex4} 

\usepackage{graphicx}

\begin{document}

\title{Constraints on the induced gravitational wave background \\ from primordial black holes}

\author{Edgar Bugaev}
\email[e-mail: ]{bugaev@pcbai10.inr.ruhep.ru}

\author{Peter Klimai}
\email[e-mail: ]{pklimai@gmail.com}
\affiliation{Institute for Nuclear Research, Russian Academy of
Sciences, 60th October Anniversary Prospect 7a, 117312 Moscow, Russia}




\begin{abstract}

We perform a consistent calculation of primordial black hole (PBH) mass spectrum and second-order
induced gravitational wave (GW) background produced from primordial scalar perturbations
in radiation era of the early Universe. It is shown that the maximal amplitudes
of the second order GW spectrum that can be approached without conflicting with the
PBH data do not depend significantly on the shape of primordial perturbation spectrum. The
constraints on the GW background obtained in previous works are extended to a wider GW
frequency range. We discuss the applicability of the currently available pulsar timing limits
for obtaining the constraints on scalar power spectrum and PBH abundance and show that
they can be used for strongly constraining the PBH number density in the PBH mass range
$\sim (0.03 - 10) M_{\odot}$.

\end{abstract}


\maketitle

\section{Introduction}
\label{sec-intro}

It is well known now that gravitational waves (GWs) can be effectively generated by
density perturbations during the radiation dominated era.
Tensor and scalar perturbations are decoupled at the first order, but it is not so
in higher orders of cosmological perturbation theory. Namely, the primordial
density perturbations and the associated scalar metric perturbations generate
a cosmological background of GWs at second order through a coupling of modes
\cite{Matarrese:1993zf, Matarrese:1997ay, Carbone:2004iv}. In particular, a second order
contribution to the tensor mode, $h_{ij}^{(2)}$, depends quadratically on the first order
scalar metric perturbation, i.e., the observed scalar spectrum sources the generation of
secondary tensor modes. By other words, the stochastic spectrum of second order GWs is induced
by the first order scalar perturbations. Calculations of $\Omega_{GW}$ at second order and
discussions on perspectives of measurements of the second order GWs are contained in works
\cite{Mollerach:2003nq, Ananda:2006af, Baumann:2007zm, Saito:2008jc, Bugaev:2009zh, Saito:2009jt}.

It is natural to conjecture that the detection of GWs from primordial density perturbations
on small scales (not directly probed by observations) could be used to constrain
overdensities on these scales, in a close analogy with the case of
primordial black holes (PBHs).
However, at the present time, gravitational wave background (GWB)
is not yet detected. So, on the contrary, one can constrain GWB using existing limits on
amplitudes of primordial density perturbations. Such limits are available, in particular,
from studies of primordial black hole production in the radiation era.

It is generally known that PBHs form from the density perturbations, induced by quantum vacuum
fluctuations during inflationary expansion. For an efficient production of PBHs in the
early Universe \cite{Zeldovich1967, Hawking1971, Carr:2005zd, Khlopov:2008qy}
the spectrum of the density perturbations set down by inflation must be ``blue'', i.e., it
must have more power on small scales. This implies that the spectral index of the scalar perturbations
must be larger than 1, in strong contradiction with the latest WMAP results
\cite{Spergel:2006hy, Dunkley:2008ie, Komatsu:2008hk}.
Such a conclusion is correct, however, only in rather special case: namely, it is
based on the prediction of slow-roll single field inflationary scenario, according to which
the power spectrum of curvature perturbations is nearly scale-invariant, i.e.,
the spectral index $n$ is close to unity and the variation in the spectral index
$dn/d \log k$ is small.

Although a prediction of the approximate scale invariance of the primordial power
spectrum is a necessary requirement to any inflationary model, some deviations from pure scale
invariance are consistent with the observational data. These deviations are described
by adding localized features to the primordial spectrum (see, e.g., \cite{Hoi:2007sf} and references
therein) and/or by introducing spectral features modifying a single power law. Models
with such peculiarities (sometimes called broken-scale-invariant (BSI) models) were
proposed, in main aspects, in eighties \cite{Starobinsky:1986fxa, Kofman:1985aw,
Kofman:1986wm, Silk:1986vc, Kofman:1988xg, Salopek:1988qh}. Such models generally
include, in addition to the usual inflaton field, other scalar fields driving
successive stages of inflation and triggering phase transitions.

Evidently, the BSI models of inflation could predict, generically, the essential production
of primordial black holes at small and medium scales. In particular, second order
phase transitions during inflationary expansion had been first considered in
\cite{Kofman:1986wm, Kofman:1988xg} in models with two scalar fields. In scenarios of
such type, during a short stage, corresponding to the beginning of a phase transition,
the mass of the trigger field becomes negative, and adiabatic perturbations are exponentially
amplified resulting in the formation of a narrow spike in the primordial spectrum, and, as
a consequence, in a copious production of PBHs \cite{GarciaBellido:1996qt, Randall:1995dj}.
There are many multiple field scenarios predicting the existence of spike- or bumplike
features in the primordial spectrum (e.g., supersymmetric double hybrid models \cite{Lesgourgues:1999uc},
multiple inflation models based on supergravity \cite{Adams:1997de}, etc). Some of these models
are specially constructed to predict efficient PBH production \cite{Yamaguchi, Kawasaki}.

An existence of the narrow spikes in the primordial spectrum is possible not only in multiple
field inflationary scenarios. Such a feature can, in principle, exist even in single field
models (see, e.g., \cite{Saito:2008em, Bugaev:2008bi}). If, e.g., the inflationary potential
has an unstable maximum at origin (e.g., the double-well potential) then, with some
fine-tuning of parameters and initial conditions, the inflation process may have two stages,
with a temporary stay at the maximum, that may lead to the corresponding peak in the
primordial spectrum and, depending on the amplitude of the peak, to the PBH production.

The details of the PBH formation from the density perturbations
had been studied in \cite{Carr:1975qj, Khlopov:1980mg}, the
astrophysical and cosmological constraints on the PBH density had been obtained in many
subsequent works (see, e.g., the recent reviews \cite{Khlopov:2008qy, Carr:2009jm}). The order of
magnitude of the corresponding constraint on the value of the density perturbation amplitude is well known
\cite{Carr:1994ar}, but, if the primordial spectrum contains the peak-like feature, the concrete
value of the PBH constraint clearly depends on the parameters characterizing the form of
this feature (in particular, on the width of the peak). Such an information may be rather
useful for the model makers.

The aim of the present work is two-fold. In the first part of the work we obtain constraints
on a power spectrum of the primordial fluctuations (for the particular case, when the spectrum has a peak
feature), for a wide range of PBH masses ($10^9 \div 10^{38}\;$g). Recently, the constraints on the
curvature perturbation from PBHs had been compiled and updated in Ref. \cite{Josan:2009qn}. Authors
of \cite{Josan:2009qn} assume that the PBHs form at a single epoch and that, over the scales probed
by a specific PBH abundance constraint, the curvature power spectrum can be written as a power law
(with a spectral index close to 1). In contrast with this, we assume that the curvature perturbation
spectrum has a peak, and the position of this peak determines the epoch of the PBH production.
A width of this peak is a model parameter, and the peak value is constrained by corresponding data
(on nucleosynthesis, photon extragalactic background, cosmological energy density parameter).

In the second part of the work we use the constraints on the curvature perturbation derived in
such a way for constraining the energy density of the induced GW background (different values
of PBH masses correspond to different values of a frequency of this background).
In our previous work \cite{Bugaev:2009kq} a part of these constraints were obtained, for rather narrow
range of frequencies ($\sim 10^{-3} \div 10^3\;$Hz), whereas in this paper we do it for the interval
$\sim 10^{-10} \div 10^4\;$Hz.

The plan of the paper is as follows. In Sec. \ref{sec-2} we present the main relations which are
necessary for a PBH mass spectrum calculation. In Sec. \ref{sec-3} we introduce our parametrization
of the power spectrum of primordial curvature perturbations, having a peak feature, and demonstrate
a dependence of the PBH mass spectrum on a width of the peak. In Sec. \ref{sec-4} we obtain the
constraints on the peak value of the primordial curvature spectrum from non-observation
of PBHs and products of their Hawking evaporation. In Sec. \ref{sec-5} we give the main formulas
used for the calculation of the induced GWB. Constraints on $\Omega_{GW}$ derived from
PBH constraints on the primordial curvature spectrum are presented in Sec. \ref{sec-6}. The
last Section contains our conclusions and discussions.


\section{PBH mass spectrum calculation}
\label{sec-2}

The calculation of PBH mass spectrum in Press-Schechter formalism
\cite{PS} is based on the expressions \cite{Kim:1996hr, Kim:1999xg, BugaevD65}
\begin{eqnarray}
n_{BH}(M_{BH}) d M_{BH} = \;\;\;\;\;\;\;\;\;\;\;\;\;\;\;\;\;\;\;\;\;\;\;\;\;\;\;\;\;\;\;\;\;\;\;\;\;\;\;\;\;\;\;\;\;
 \label{nBHMBH}\\
= \left \{ \int n(M, \delta_R) \frac{d\delta_R}{d\delta_R^H} \frac{dM}{dM_{BH}}
d\delta_R^H \right \}dM_{BH} \nonumber \;,
\\
\label{nBHMBH2}
n(M, \delta_R) = \sqrt{\frac{2}{\pi}} \frac{\rho_i}{M}
\frac{1}{\sigma_R^2} \left| \frac{\partial \sigma_R}{\partial M}
\left( \frac{\delta_R^2}{\sigma_R^2} -1 \right) \right|
e^{-\frac{\delta_R^2}{2\sigma_R^2}}.
\end{eqnarray}

Here, the following notations are used: $\delta_R$ is the initial (at the moment $t_i$)
density contrast smoothed on the comoving scale $R$, $M$ is the smoothing
mass (initial mass of the fluctuation
corresponding to the scale $R$), $\sigma_R(M)$ is the mean square deviation (the mass variance),
\begin{equation}
\sigma_R^2(M) = \int \limits_{0}^{\infty} {\cal P}_\delta(k) W^2(kR)
\frac{dk}{k}, \label{SigR}
\end{equation}
${\cal P}_\delta(k)$ is the power spectrum of primordial density
perturbations, $W(kR)$ is the Fourier transform of the window function (in this work we use
the gaussian one, $W(kR) = \exp (- k^2 R^2/2 )$ ),
$\rho_i$ is the initial energy density.
It is assumed that the process of reheating is very short in time, so, the end
of inflation practically coincides with a start (at $t=t_i$) of the radiation era.

Fourier transform of the (comoving) density contrast is
\begin{equation}
\delta_k(t) = -\frac{2}{3} \left(\frac{k}{aH} \right)^2 \Psi_k(t),
\label{dkt}
\end{equation}
where $\Psi_k(t)$ is the Fourier transform of the Bardeen potential.
Here, we explicitly take into account the time dependence
of the Bardeen potential.

The power spectrum of the density perturbations, calculated at some moment of time, is
\begin{equation} \label{Pdeltak}
{\cal P}_\delta(k, t) = \left[ \frac{2}{3} (k \tau)^2  \right]^2 {\cal P}_{\Psi}(k, t),
\end{equation}
where $\tau$ is the conformal time ($\tau=(aH)^{-1}$ for the radiation
epoch).

The comoving smoothing scale, $R\equiv 1/k_{R}$, is connected with the smoothing mass $M$ by the expression
\begin{equation}
\label{kflaiHi} %
\Big( \frac{M}{M_i} \Big) ^{-2/3} = \frac{k_{R}^2}{(a_i H_i)^2},
\end{equation}
where $M_i$, $a_i$ and $H_i$ are the horizon mass, cosmic scale factor and Hubble parameter at the
moment $t_i$.

In the approximation of instantaneous transition from inflationary era to the radiation
dominated epoch, the connection between density perturbation at any time and curvature perturbation at
initial moment of time $t_i$ is \cite{Bugaev:2008gw}
\begin{equation} \label{Pdeltak2}
{\cal P}_\delta(k, t) = \left[ \frac{2}{3} (k \tau)^2
\frac{\Psi_k(\tau)}{ {\cal R}_k(\tau_i)} \right]^2 {\cal P}_{\cal R}(k, t_i),
\end{equation}
where the expression for $\Psi_k(\tau)$ is given by \cite{Lyth:2005ze}
\begin{eqnarray}
\label{PsiSol} \Psi_k(\tau) = \frac{2 {\cal R}_k(\tau_i)}{x^3} [
(x-x_i) \cos(x-x_i) -  \\ \nonumber -
(1+xx_i)\sin(x-x_i) ] , \\ \nonumber
x = \frac{k\tau}{\sqrt{3}}, \;\; x_i = \frac{k\tau_i}{\sqrt{3}}.
\end{eqnarray}
Here, ${\cal R}_k(\tau_i)$ is the Fourier component of the curvature perturbation on
the comoving hypersurfaces, at the end of inflation (see, e.g., \cite{Lyth:2005ze}).

The connection between values of the smoothing mass $M$, density contrast $\delta_R^H$
and PBH mass $M_{BH}$ can be expressed in the general form
\begin{equation}
M_{BH} = \varphi(M, \delta_R^H; M_i) .
\end{equation}
The concrete expression for the function $\varphi$ depends on the
model of the gravitational collapse. In the model of the standard
spherically-symmetric collapse the connection is quite simple:
\begin{equation}
\label{MBHsph} %
M_{BH} =  (\delta_R^H)^{1/2} M_h .
\end{equation}
Here, $M_h$ is the horizon mass at the moment of time, $t=t_h$,
when regions of the comoving size $R$ and smoothing mass $M$ cross horizon.
According to Carr and Hawking \cite{CarrHawking}, $1/3 = \delta_{th} \le \delta_R^H \le 1$. The derivation
of Eq. (\ref{MBHsph}) is given in the Appendix of \cite{Bugaev:2008gw}.
From (\ref{MBHsph}), using the relation $M_h = M_i^{1/3} M^{2/3}$, one has the expression for the
function $\varphi$ for the Carr-Hawking collapse:
\begin{equation}
\label{fCH} %
\varphi(M, \delta_R^H; M_i) = (\delta_R^H)^{1/2} M^{2/3} M_i^{1/3}.
\end{equation}

In the picture of the critical collapse \cite{NJ, Musco:2008hv} the corresponding function is
\begin{equation}
\label{fcrit} %
\varphi(M, \delta_R^H; M_i) = k_c (\delta_R^H-\delta_c)^{\gamma_c}
M^{2/3} M_i^{1/3},
\end{equation}
where $\delta_c$, $\gamma_c$ and $k_c$ are model parameters.
The mass spectrum of PBHs for the critical collapse model has been calculated, e.g., in \cite{Bugaev:2008gw}.
It was shown that for the primordial scalar perturbation spectrum with a peak
the maximum of PBH mass spectrum is still around the horizon mass corresponding
to the maximum in primordial power spectrum, but the PBH mass spectrum also has a ``tail'' of
small masses. In this work we will use classical collapse model (Eq. (\ref{fCH})).
However, in the end of the paper we will explore the dependence of the results on $\delta_{th}$ - the density
contrast threshold of PBH formation, which gives the main uncertainty due to
the exponential sensitivity of PBH abundance to it.

\section{Primordial power spectrum with maximum}
\label{sec-3}

\begin{figure}
\includegraphics[trim = 0 0 0 0, width=7.8 cm]{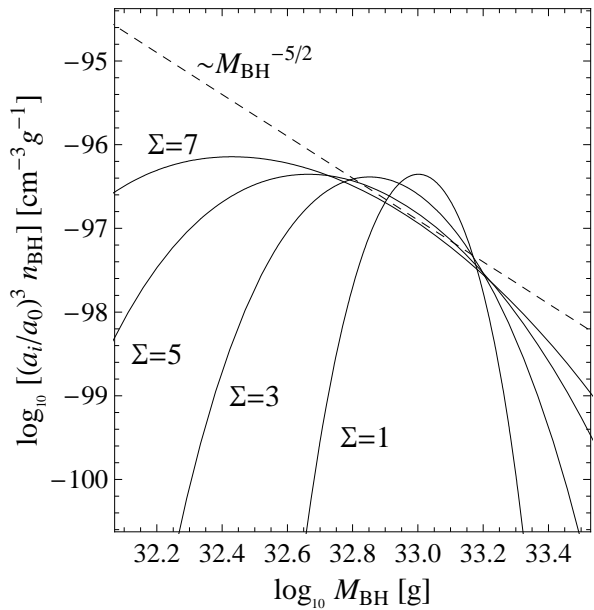} %
\caption{The calculation of PBH mass spectra for a peaked ${\cal P}_{\cal R}(k)$-spectrum,
for $M_h^0=10^{33}\;$g and different values of $\Sigma$. The value of ${\cal P}_{\cal R}^0$ was chosen
so that for all cases $\Omega_{PBH} \approx 0.024$
(for $\Sigma=1$, ${\cal P}_{\cal R}^0=0.0488$;
for $\Sigma=3$, ${\cal P}_{\cal R}^0=0.0294$;
For $\Sigma=5$, ${\cal P}_{\cal R}^0=0.0267$;
For $\Sigma=7$, ${\cal P}_{\cal R}^0=0.0258$.)
The dashed curve shows, for comparison, the
form of the classical Carr's $M_{BH}^{-5/2}$-mass spectrum \cite{Carr:1975qj}.}
\label{fig-pbh-sp}
\end{figure}

It is convenient to use some kind of parametrization to model the realistic peaked power
spectrum of finite width. We use the distribution of the form
\begin{equation}
\label{PRparam} %
\lg {\cal P}_{\cal R} (k) = B + (\lg {\cal P}_{\cal R}^0 - B)
\exp \Big[-\frac{(\lg k/k_0)^2}{2 \Sigma^2} \Big].
\end{equation}
Here, $B \approx -8.6$, ${\cal P}_{\cal R}^0$ characterizes the height of the peak,
$k_0$ is the position of the maximum and $\Sigma$ is the peak's width. Parameters of such
a distribution have been constrained in the previous work of authors \cite{Bugaev:2008gw} from non-observation
of PBHs and products of their Hawking evaporation (photons and neutrinos).

In many cases limits on PBH abundance have been obtained using the assumption
that PBH mass distribution is close to the $\delta$-function form. This does not, however, mean
that the original spectrum ${\cal P}_{\cal R} (k)$ should be very narrow. For example, in the
calculation of PBH mass spectra given in Fig. \ref{fig-pbh-sp}, all distributions
have a distinct maximum (even for rather wide peak, $\Sigma=7$), which shifts
to smaller $M_{BH}$ with the growth of $\Sigma$. In this sense, if our goal was
just to explore the PBH abundance, we could assume that ${\cal P}_{\cal R} (k)\sim \delta(k-k_0)$ from
the start, without introducing big mistakes (note, however, that with such an approach
the shift of the distribution maximum would not be noticed).
This is, however, not an adequate approximation for our study, because we are also interested
in calculation of the 2-nd order GWB produced by the same power spectrum and studying
of its dependence on the width of the peak.

One should note that in Fig. \ref{fig-pbh-sp} we show on the vertical axis the quantity $n_{BH}\times (a_i/a_0)^3$,
which is the physical (rather than comoving) number density of PBHs, and is independent
on the reheating temperature $T_{RH}$ in the limit $k_{i} \equiv a_i H_i \gg k_0 = k(M_h^0)$
\cite{Bugaev:2008gw}.

The fraction of an energy density of the Universe contained in PBHs today, $\Omega_{\rm PBH}$,
assuming that a mass of the produced black hole does not change in time, is
\begin{equation}
\label{OmegaPBH}
\Omega_{PBH} = \frac{1}{\rho_c} \left(
\frac{a_i}{a_0}\right)^3 \int M_{BH} n_{BH}(M_{BH}) d M_{BH}
\end{equation}
($\rho_c$ is the critical density). This formula is rather accurate for black holes with
initial mass $M_{BH} \gg M_*$, where $M_* \approx (3 t_0 \alpha_0)^{1/3} \approx 5\times 10^{14}$ g
is the initial mass of PBH which reaches its final state of evaporation
today \cite{Page:1976df}, $\alpha_0=8.42\times 10^{25}\; {\rm g}^3 {\rm s}^{-1}$,
and $t_0$ is the age of the Universe.

\section{Constraints on ${\cal P}_{\cal R}^0$}
\label{sec-4}

For constraining ${\cal P}_{\cal R}^0$ we use the existing limits on PBH abundance.
In the region of $M_{BH}$ which is of interest for us ($10^9 \lesssim M_{BH} \lesssim 10^{38}\;$g)
these limits can be divided in three groups: {\it i)} constraints on PBHs from big bang
nucleosynthesis (due to hadron injections by PBHs \cite{Zeldovich1977}, photodissociation
of deuterium \cite{Lindley1980} and light nuclei, fragmentations of quarks and gluons
evaporated by PBHs \cite{Kohri:1999ex}), $10^9 \lesssim M_{BH} \lesssim 10^{13}\;$g, and from
influence of PBH evaporations on the CMB anisotropy,
$2.5\times 10^{13} \lesssim M_{BH} \lesssim 2.5 \times 10^{14}\;$g \cite{Carr:2009jm},
{\it ii)} constraints on PBHs from extragalactic photon background, $10^{13} \lesssim M_{BH} \lesssim 10^{17}\;$g,
{\it iii)} constraints on non-evaporating PBHs (gravitational and lensing constraints).
Constraints on PBHs from data on extragalactic neutrino
background \cite{BugaevD65, BugaevD66, Bugaev:2008gw}, in the region
$10^{11} \lesssim M_{BH} \lesssim 10^{13}\;$g, are somewhat weaker than nucleosynthesis constraints.

We do not consider in this paper the PBH constraints from PBH masses smaller than $10^9\;$g
because the corresponding GWB frequencies are too high ($\gtrsim 10^4\;$Hz).
Also, we did not consider in this paper the constraints from galactic gamma rays
(they could be essential in the narrow mass region near $M_{BH}\sim 10^{15}\;$g, but
strongly depend on the unknown clustering factor \cite{Carr:2009jm}) and potential
constraints \cite{Carr:2009jm} from future measurements of 21 cm line.

For a derivation of the constraint on ${\cal P}_{\cal R}^0$ in the region $10^9 \lesssim M_{BH} \lesssim 10^{13}\;$g
we use the latest update of the nucleosynthesis constraints given in the review \cite{Carr:2009jm}.
In this region of PBH masses we, following the traditional practice, approximate the initial PBH mass
spectrum by $\delta$-function (i.e., we assume that all PBHs have the same mass $M_{BH}$).
The mass of the PBH formed is approximately equal to the horizon mass at
horizon entry, $M_{BH}\approx M_h$. Correspondingly,
the constraints are expressed in terms of the function $\beta(M_{BH})$ which is the mass fraction
of the energy density of the Universe going to PBHs. It is given by the relation
\begin{eqnarray}
\beta(M_{BH}) = 2 \int \limits_{\delta_{th}}^{1} P(\delta_{hor}(R)) d\delta_{hor}(R) =\\
\nonumber =  {\rm erfc}\left( \frac{\delta_{th}}{\sqrt{2} \sigma_{R}(M_{BH})} \right),
\end{eqnarray}
where the mass variance is given by \cite{Bugaev:2008gw, Josan:2009qn}
\begin{eqnarray}
\sigma_{R}^2(M_{BH}) = \frac{16}{3} \int \limits_{0}^\infty (kR)^2 j_1^2\left(\frac{kR}{\sqrt{3}}\right)\\
\nonumber \times \exp \left( -k^2 R^2 \right) {\cal P}_{\cal R}(k) \frac{dk}{k} ,
\end{eqnarray}
$R$ is the smoothing scale, i.e., the horizon size.

The order of magnitude of the constraint on $\beta$ is, according to \cite{Carr:2009jm}:
$\beta \lesssim 10^{-18}$ for
$10^9{\rm g} < M_{BH} < 10^{10} {\rm g}$ and $\beta \lesssim 10^{-23}$ for
$10^{10}{\rm g} < M_{BH} < 10^{13} {\rm g}$. In the narrow region near $\sim 3\times 10^{13}\;$g
there is the strong constraint following from CMB anisotropy damping,
$\beta \lesssim 10^{-28}$.

For a derivation of the constraint on ${\cal P}_{\cal R}^0$ in the region
$10^{13} \lesssim M_{BH} \lesssim 10^{17}\;$g
we use the extended PBH mass spectrum given in Sec. \ref{sec-2} and the technique developed by
authors in \cite{Bugaev:2008gw} (in particular, the condition that photons
(primary as well as secondary ones \cite{MacGibbon:1990zk}) evaporated from
PBHs do not exceed the observed extragalactic gamma ray background was used).
Our method is essentially the same as used in the pioneering
work \cite{Page:1976wx}. Extragalactic photon background data which we used
had been obtained in \cite{Strong:2004ry} (EGRET collaboration) and in
\cite{Abdo:2010nz} (Fermi LAT  collaboration). Comparison shows that in the
energy region $\sim 0.1 \div 1\;$GeV (which is of interest for us) data
of both works are consistent with each other. So, for concrete calculations
we used the data from \cite{Strong:2004ry}.

At last, the gravitational constraint is just a condition
that $\Omega_{PBH}$ does not exceed the energy density of non-baryonic
dark matter (in the limiting case, PBHs account for all dark matter):
\begin{equation}
\label{OmegaPBHlimit}
\Omega_{PBH} \le \Omega_{CDM} \approx 0.25.
\end{equation}
The value of ${\cal P}_{\cal R}^0$ is obtained from (\ref{OmegaPBH}) and (\ref{OmegaPBHlimit}).

Lensing constraints arise due to the fact that microlensing observations of stars in the
Magellanic Clouds probe the fraction of the Galactic halo
in massive compact halo objects (MACHOs) of sub-solar masses \cite{Paczynski:1985jf}.
The fraction of the halo in PBHs is
\begin{equation}
f(M_{BH}) = \frac {\Omega_{PBH}}{\Omega_{CDM}}.
\end{equation}
From the analysis of MACHO \cite{Alcock:1998fx} and EROS \cite{Tisserand:2006zx} microlensing surveys,
$f(M_{BH}) < 0.1$ for $10^{-6} M_\odot < M_{BH} < M_\odot$ and
$f(M_{BH}) < 0.04$ for $10^{-3} M_\odot < M_{BH} < 0.1 M_\odot$.
There are some additional constraints on $f(M_{BH})$ for other mass ranges, reviewed, e.g.,
in \cite{Carr:2009jm}, but they are generally rather weak ($f(M_{BH}) \approx 1$) and we do
not use them here.

\begin{figure}
\includegraphics[trim = 0 0 0 0, width=8.5 cm]{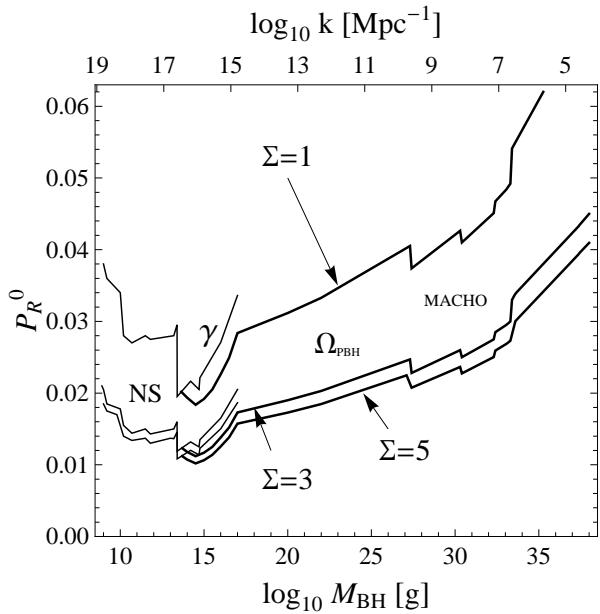} %
\caption{
The limits on the value of ${\cal P}_{\cal R}^0$ from PBH non-observation,
for three different values of $\Sigma$. The constraints based on the $\beta$-function
approach are shown by thin lines.}
\label{fig-PR0-bound}
\end{figure}

The resulting constraints are shown in Fig. \ref{fig-PR0-bound}.
The constraints based on the $\beta(M_{BH})$ function are designated by thin lines
(the corresponding constraints on $\beta$ have been taken from Fig. 6 of \cite{Carr:2009jm}).
The constraints obtained by using our technique (which is briefly explained is Sec. \ref{sec-2})
and EGRET data, and gravitational constraints are shown by thick lines. For
calculations of limits on $\Omega_{GW}$ in the following Sections
we use, for smaller masses (before
the crossing points of thin and thick lines), the ``$\beta$-constraint'', and for
larger masses, after the crossing points, the constraints given by the thick lines.
The value of the crossing point is about $4\times 10^{13}\;$g, almost independently
on the value of $\Sigma$.

One can see from Fig. \ref{fig-PR0-bound} that the results for ${\cal P}_{\cal R}^0$
constraint depend on the method of the calculation, in the region $\sim 10^{14} \div 10^{17}\;$g.
Namely, the constraints based on the $\beta$ function are somewhat weaker. We did
not study, in the present work, the dependence of this result on the form of the
PBH mass spectrum.

The ${\cal P}_{\cal R}^0$-constraints shown in
Fig. \ref{fig-PR0-bound} (for the cases of $\Sigma=3,5$)
are close to the ones derived in \cite{Josan:2009qn}. Authors of \cite{Josan:2009qn}
have assumed that the power spectrum is scale-invariant over the
(relatively small) range of scales which contribute to a given
constraint (this is analogous to the assumption that
primordial spectrum has a rather wide peak).
They argued that deviations from scale-invariance (which are consistent with
a slow-roll inflation hypothesis) lead to small changes in the constraints.
However, in the case of a narrow peak (our case of $\Sigma=1$), which can be produced,
e.g., by a violent slow-roll violation during single-field inflation,
the limit on ${\cal P}_{\cal R}^0$ will, of course, be much
weaker, as we see from Fig. \ref{fig-PR0-bound} (because the total power contained in the spectrum
depends on its width).

\section{Induced GW background}
\label{sec-5}

\subsection{Basic formulas}

According to \cite{Baumann:2007zm}, the power spectrum of induced GWs is given by the expression
\begin{equation} \label{Phktau}
{\cal P}_h(k, \tau) = \int\limits_0^\infty d\tilde k \int \limits_{-1}^{1} d\mu  \;
{\cal P}_\Psi(|{\bf k-\tilde k|}) {\cal P}_\Psi (\tilde k) {\cal F}(k,\tilde k,\mu, \tau),
\end{equation}
where
\begin{eqnarray}
\label{calF}
{\cal F}(k,\tilde k,\mu, \tau) = \frac{(1-\mu^2)^2}{a^2(\tau)} \frac{k^3 \tilde k^3}{|{\bf k-\tilde k}|^3}
\times \nonumber \;\;\;\;\;\;\;\;\;\;\;\;\;\;\;\;\;\;\;\;\;\;\;\;\;\;\;\;\;\; \\  \times
\int \limits_{\tau_0}^{\tau} d \tilde \tau_1 \; a(\tilde \tau_1) g_k(\tau, \tilde\tau_1)
f({\bf k},{\bf \tilde k}, \tilde \tau_1) \times \;\;\;\;\;\;\;\;\;\;\;\;\;\;\;\;\;\;\;\;\; \\ \times
\int \limits_{\tau_0}^{\tau} d \tilde \tau_2 \; a(\tilde \tau_2) g_k(\tau, \tilde\tau_2)
\left[f({\bf k},{\bf \tilde k}, \tilde \tau_2) + f({\bf k},{\bf k-\tilde k}, \tilde \tau_2)\right] \nonumber
\end{eqnarray}
and
\begin{eqnarray} \label{f}
f({\bf k},{\bf \tilde k}, \tau) = 12 \Psi(\tilde k\tau) \Psi(|{\bf k-\tilde k}| \tau) +
 \;\;\;\;\;\;\;\;\;\;\;\;\;\;\;\;\;\;\;\;\; \;\;\;\;\;\; \\ \nonumber
+ 8\tau \Psi(\tilde k\tau)
\Psi'(|{\bf k-\tilde k}| \tau) + 4 \tau^2 \Psi'(\tilde k\tau)\Psi'(|{\bf k-\tilde k}| \tau).
\end{eqnarray}
In Eqs. (\ref{Phktau}, \ref{calF}, \ref{f}) the following notations are used. ${\cal P}_\Psi(k)$ is
the power spectrum of the Bardeen potential, defined at some moment of time
$\tau=\tau_i'$ near the beginning of the RD stage
(by definition, it is the primordial spectrum),
\begin{equation} \label{Ppsilr}
\langle \Psi_{\bf k} \Psi_{\bf k'} \rangle = \frac{2\pi^2}{k^3} \delta^3({\bf k}+{\bf k'}) {\cal P}_\Psi(k),
\end{equation}
$\Psi_{\bf k}$ is the Fourier component of $\Psi$,
\begin{equation}
\Psi({\bf x}) = \frac{1}{(2\pi)^{3/2}} \int d^3 {\bf k} \Psi_{\bf k} e^{i{\bf k}\cdot{\bf x}},
\end{equation}
$\mu = {\bf k \cdot \tilde k} / (k \tilde k)$ is the cosine of the angle between the vectors
${\bf k}$ and ${\bf \tilde k} $. The power spectrum of GWs is defined by the standard expression
\begin{equation}
\langle h_{\bf k}(\tau) h_{\bf k'}(\tau) \rangle = \frac{1}{2} \frac{2\pi^2}{k^3}
\delta^3({\bf k}+{\bf k'}) {\cal P}_h(k, \tau),
\end{equation}
where $h_{\bf k}(\tau)$ is the Fourier component of the tensor metric perturbation,
\begin{equation}
h_{ij}(x, \tau) = \int \frac{d^3 {\bf k}}{(2\pi)^{3/2}} e^{i{\bf k}\cdot{\bf x}}
\left[ h_{\bf k}(\tau) e_{ij}({\bf k}) + \bar h_{\bf k}(\tau) \bar e_{ij}({\bf k})\right],
\end{equation}
$e_{ij}({\bf k})$ and $\bar e_{ij}({\bf k})$ are two polarization tensors corresponding to the
wave number ${\bf k}$.

The function $f$ in Eq. (\ref{f}) contains transfer functions $\Psi(k\tau)$, which are defined by
\begin{equation} \label{PsiTRFUN}
\Psi(k \tau) = \frac{\Psi_k(\tau)}{\Psi_k},
\end{equation}
where $\Psi_k \equiv \Psi_k(\tau_i')$ is the initial (primordial) value of the potential.
During RD epoch, the solution for the Bardeen potential, having the initial condition
$\Psi_k(\tau_i)=0$, where $\tau_i$ is the moment of the end of inflation which is close
to $\tau_i'$ (but $\tau_i<\tau_i'$), is given by Eq. (\ref{PsiSol}).
The value of the potential at $\tau_i$ is chosen to be zero because $\Psi_k$ is
typically very small during inflation \cite{Lyth:2005ze} and it is a continuous function during
the transition from inflationary to RD stage (we assume, for simplicity, that the reheating
is instant).
We have chosen, for the numerical calculation, the value of $\tau_i'$ using the
condition $\lg(\tau_i'/\tau_i)=0.05$ and thereby neglect the formation of PBHs and induced GWs in the
interval of time from $\tau_i$ to $\tau_i'$. In this work we are interested only in
wave numbers $k\ll k_{end}$, for which there is no dependence of the results on $\tau_i$
(because the perturbation amplitudes, such as $\Psi_k(\tau)$, have enough time to
reach their asymptotic limit before horizon re-entry for each mode).
For an example of the case where values of $k\sim k_{end}$ are important, see, e.g., Ref. \cite{Bugaev:2008gw}
(where the case of the running mass model is considered).

The function $g_k(\tau, \tilde\tau)$ in Eq. (\ref{calF}) is the Green function
which depends on the cosmological epoch. For RD Universe,
\begin{equation}
g_k(\tau, \tilde \tau) = \frac{1}{k} \sin[k(\tau - \tilde\tau)] \;\;\; , \;\;\; \tau < \tau_{\rm eq}.
\label{gk-RD}
\end{equation}

The energy density of GWs per logarithmic interval of $k$ in units of the critical density is given by
\begin{equation} \label{OmegaGW}
\Omega_{GW}(k, \tau) =
\frac{1}{12} \left( \frac{k}{a(\tau) H(\tau)} \right)^2 {\cal P}_h(k, \tau).
\end{equation}
Here, the power spectrum of GWs, ${\cal P}_h(k, \tau)$, is obtained from the
formula (\ref{Phktau}). However, for very large
wave numbers $k$ which we are interested in, the direct use of (\ref{Phktau}) will require
numerical integration
for functions having a huge number of oscillations (e.g., for $k \sim 10^{16} {\rm Mpc}^{-1}$
this is about $\sim k \tau_0 \sim 10^{20}$ oscillations). This is hard to do numerically.
Fortunately, we do not have to do integration until the present day. As discussed in \cite{Bugaev:2009zh},
it is enough
to calculate $\Omega_{GW}$ for the moment of time $\tau_{calc} \gg k^{-1}$ at which
the mode is well inside the horizon, and is freely propagating. We can then easily relate
energy densities of GWs at different times with simple calculation, using the fact that
$h_k\sim a^{-1}$ far inside the horizon
(this approach works well only for rather large values of $k$,
$k \gtrsim k_c \approx 100 k_{eq} \approx 1 {\rm Mpc}^{-1}$, but we are only interested
in such large wave numbers here).

The final expression for $\Omega_{GW}^0(k)$ is \cite{Bugaev:2009zh}
\begin{equation}
\Omega_{GW}^0(k) =
2 \Omega_R  \left( \frac{g_{* eq}}{g_{* calc}}\right)^{1/3} \times \frac{(k\tau_{\rm calc})^2}{12}
{\cal P}_h(k, \tau_{\rm calc}).
\end{equation}
This formula gives the correct energy density, accurate to the oscillations in it.
In practice, $\tau_{\rm calc}$ can be either fixed or
dependent on $k$, e.g., for the last case,
\begin{equation}
\tau_{\rm calc} = N_{\rm sub} \cdot k^{-1}, \;\;\;\; N_{\rm sub} \sim 100.
\end{equation}
It proves to be more convenient to use the ``randomized'' value of $N_{\rm sub}$, i.e.,
\begin{equation}
\tau_{\rm calc} = (\tilde N_{\rm sub}+N_{\rm rnd}) \cdot k^{-1},
\label{Nk}
\end{equation}
where $\tilde N_{\rm sub}$ is constant and $N_{\rm rnd}$ is a random number in the
interval $[0, 2\pi]$ calculated independently for
every $k$. In this case the result of the calculation is a stochastically oscillating function
whose envelope always can be easily found, and it is the envelope that we are interested in.
The exact shape of
the function will, actually, depend on the choice of $\tau_{\rm calc} \gg k^{-1}$, and the
larger $\tau_{\rm calc}$ we take, the more frequent are the oscillations, but the envelope which
we are interested in does not change.
This was explicitly shown in work \cite{Bugaev:2009zh} (see, in particular, Figure 2 from it,
which shows the same GW spectrum calculated using approach of Eq. (\ref{Nk})
and the one using $\tau_{calc}={\rm const}$. It is seen from that figure that the resulting
spectrum is the same).

\subsection{Connection between frequency and horizon mass}

For a wave with comoving wave number $k$ and wavelength $\lambda=2\pi/k$, propagating at the speed of light
$c$, the corresponding frequency is $f=c/\lambda $, or
\begin{equation}
f=\frac{ck}{2\pi} = 1.54\times 10^{-15 } \left( \frac{k}{{\rm Mpc}^{-1}} \right) {\rm Hz}.
\label{fck}
\end{equation}
From the constancy of the entropy in the comoving volume, we have the relation between the scale factor $a$,
temperature $T$ and the effective number of degrees of freedom $g_*$:
\begin{equation}
a\sim g_*^{-1/3} T^{-1}.
\end{equation}
From the Friedmann equation ($H^2 \sim \rho$), we have
\begin{equation}
H \sim a^{-2} g_*^{-1/6}, \label{Hag}
\end{equation}
and the horizon mass corresponding to the scale factor $a$ evolves during the RD
epoch as
\begin{equation}
M_h \sim (H^{-1})^3 \rho \sim a^2 g_*^{1/6}. \label{Mhag}
\end{equation}
From (\ref{Hag}) and (\ref{Mhag}), the wave number of the mode entering horizon at the moment of
time $t$ (at this time, $k=aH$) is related to the horizon mass at the same moment of time by
\begin{eqnarray}
k = k_{eq} \left( \frac{M_h}{M_{eq}} \right)^{-1/2} \left( \frac{g_*}{g_{* eq}} \right)^{-1/12}
\approx \nonumber \\
\;\;\;\;\;\;\;\;\;\;\; \approx 2 \times 10^{23} (M_h[{\rm g}])^{-1/2} \;\; {\rm Mpc}^{-1},
\label{kkeq}
\end{eqnarray}
where in the last equality we have adopted that $g_{* eq} \approx 3$, $g_* \approx 100$,
\begin{equation}
M_{eq} = 1.3 \times 10^{49} {\rm g} \cdot (\Omega_m h^2)^{-2} \approx 8 \times 10^{50} {\rm g},
\end{equation}
\begin{equation} \label{keq}
k_{eq} = a_{eq} H_{eq} =  \sqrt{2} H_0 \Omega_m \Omega_R^{-1/2} \approx 0.0095 \; {\rm Mpc}^{-1}.
\end{equation}

The factor $g_*$ is a function of cosmic temperature $T$ and in this work we assume that
there are no degrees of freedom beyond the Standard Model, so $g_*\approx 100$ for $T \gtrsim 100\;$MeV,
$g_* \approx 10$ for $100 \; {\rm MeV} \gtrsim T \gtrsim 1 \; {\rm MeV}$ and
$g_* \approx g_{* eq} \approx 3$ for $T \lesssim 1\;$ MeV. The connection between horizon mass
and $T$ is
\begin{equation}
M_h \approx 3 \times 10^5 M_\odot
\left( \frac{g_*}{g_{* eq}} \right)^{-1/2}
\left( \frac{T}{1 \; {\rm MeV}} \right)^{-2},
\end{equation}
and we can estimate that $M_h(1 \; {\rm MeV}) \approx 10^5 M_\odot$, and
$M_h(100 \; {\rm MeV}) \approx 5 M_\odot$.

The frequency of the wave corresponding to the wave number $k$ can be related to the horizon mass
by the relation following from (\ref{fck}) and (\ref{kkeq}),
\begin{equation}
f \approx 3 \times 10^8 \;{\rm Hz} \times (M_h[{\rm g}])^{-1/2} ; \;\;\;
M_h \approx \frac{9 \times 10^{16} {\rm g}}{(f[{\rm Hz}])^2}.
\label{fMh}
\end{equation}
For scalar-induced GWs, the single mode in scalar spectrum does not correspond to the only
one mode in ${\cal P}_h$. For example, for the $\delta$-function-like spectrum
${\cal P}_{\cal R}(k) \sim \delta(k-k_0)$, the GW spectrum is continuous and stretches from $0$
to $2k_0$ \cite{Ananda:2006af}. However, the order of magnitude of wave numbers of induced GWs
is the same as of scalar perturbations, so (\ref{fMh}) gives an estimate of GW frequency
that will be generated from perturbations entering horizon at its mass scale $M_h$. Furthermore, if
PBHs form from a scalar spectrum of perturbations at a horizon mass scale $M_h$, the typical
PBH mass will be of order of $M_h$, so (\ref{fMh})
relates the typical PBH mass with the characteristic frequency of second-order GWs produced.

\section{Constraints on $\Omega_{GW}$ from PBHs}
\label{sec-6}

\begin{figure}
\includegraphics[trim = 0 0 0 0, width=8.4 cm]{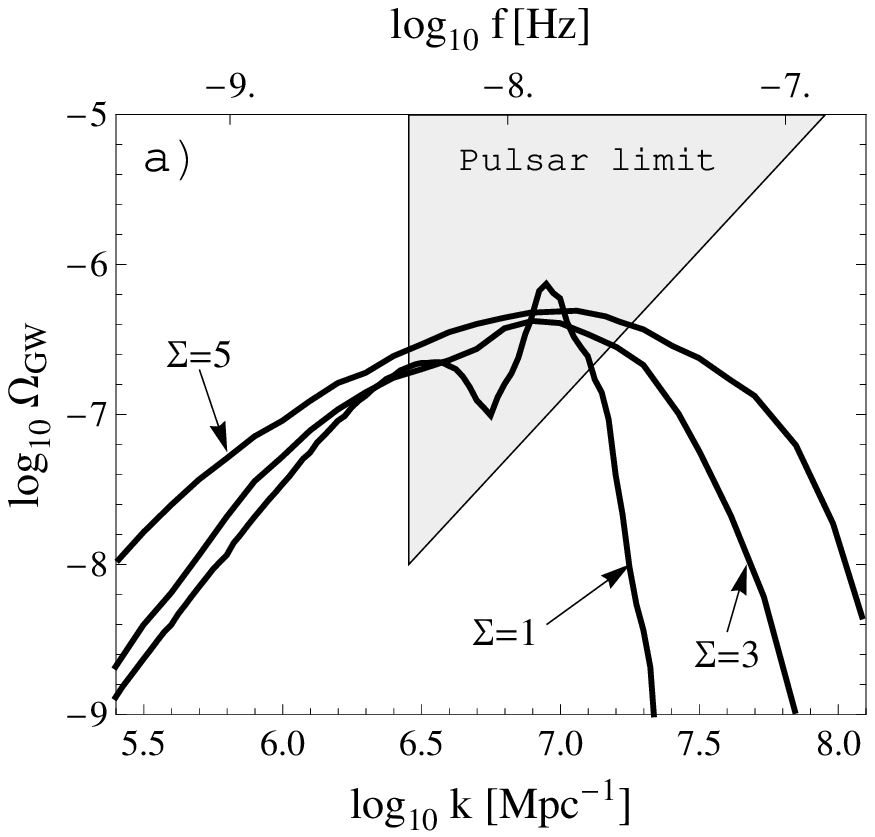} \\ $\;$ \\ $\;$
\includegraphics[trim = 0 0 0 0, width=8.4 cm]{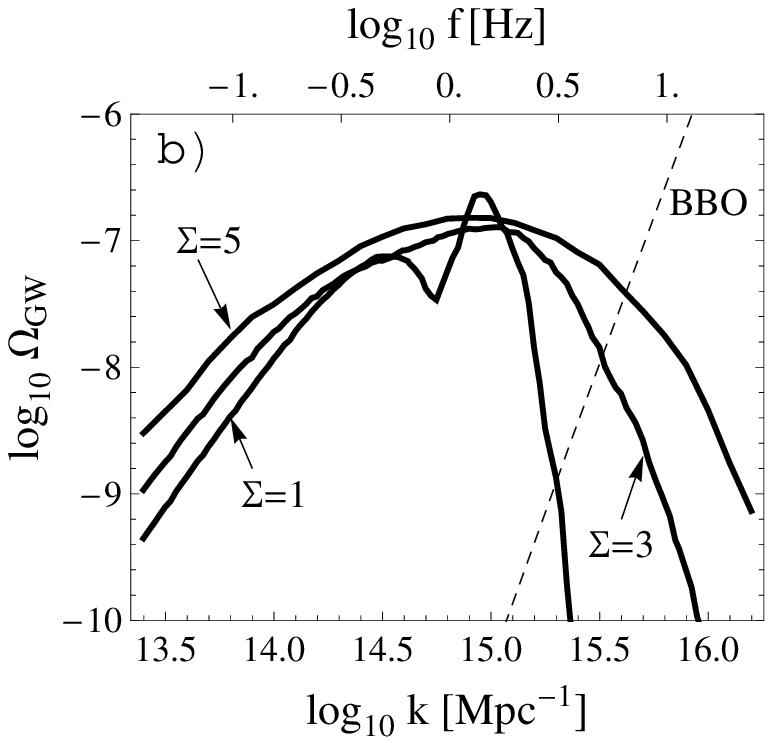}
\caption{
{\bf a)} The calculation of the induced GWB corresponding to primordial power spectra and
PBH mass spectra given in Fig. \ref{fig-pbh-sp} (cases of $\Sigma=1, 3, 5$).
The pulsar timing limit for $\Omega_{GW}$ is also shown.
{\bf b)} The calculation of the induced GWB for $M_h^0=10^{17}\;$g
(for all three cases, $\Omega_{PBH}\approx 0.24$, the corresponding sets of parameters are
${\cal P}_{\cal R}^0=0.028$ for $\Sigma=1$,
${\cal P}_{\cal R}^0=0.0172$ for $\Sigma=3$, and
${\cal P}_{\cal R}^0=0.0159$ for $\Sigma=5$).
}
\label{fig-gw-diff}
\end{figure}

In Fig. \ref{fig-gw-diff} we show the result of $\Omega_{GW}$ calculation for the finite-width
distribution of the form (\ref{PRparam}). It is seen that for a narrow peak, the distribution
looks much like the double-peaked one produced by a $\delta$-function power spectrum
\cite{Ananda:2006af, Bugaev:2009zh}. The shape is smoothing
with the growth of $\Sigma$ (and it will be a scale-invariant spectrum for the scale-invariant
input \cite{Baumann:2007zm}). The value of $\Omega_{GW}$ in the case of rather wide
peak ($\Sigma \gg 1$) is proportional
to $({\cal P}_{\cal R})^2$ and can be estimated as
\begin{equation}
\label{OmegaApprox2}
\Omega_{GW}(k>k_c, \tau_0)\cong 0.002 \left( \frac{g_{* eq}}{g_*} \right)^{1/3} \cdot {\cal P}_{\cal R}^2 .
\end{equation}

\begin{figure}
\includegraphics[trim = 0 0 0 0, width=8.8 cm]{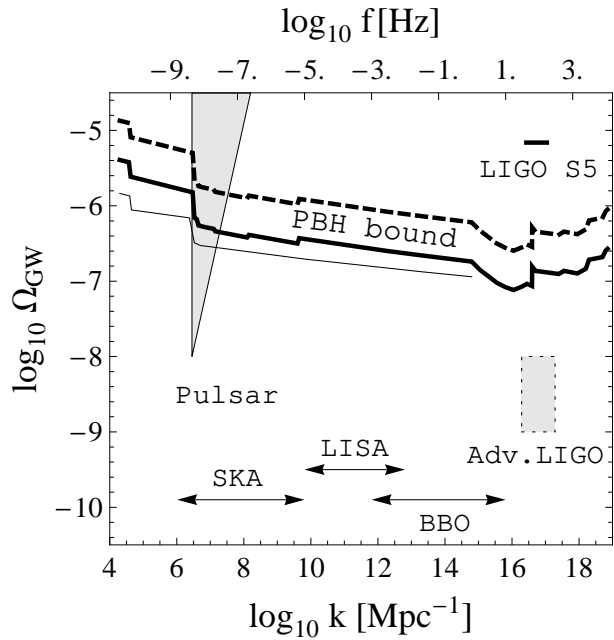} %
\caption{
The limits on the 2-nd order GWB from primordial black holes, obtained in this paper.
Solid thin line shows the result assuming PBH formation threshold is $\delta_{th}=1/3$.
Dashed line corresponds to the assumption $\delta_{th}=0.45$. Thin line
shows the maximum values of GWB that can be reached for $\Omega_{PBH}=10^{-5}$ (and $\delta_{th}=1/3$).
Also shown are current pulsar timing limit, LIGO S5 limit, Advanced LIGO planned
sensitivity to $\Omega_{GW}$, and frequency ranges in which other future experiments
(LISA, BBO, SKA) will operate. The estimate sensitivity of all future experiments,
in the corresponding frequency ranges, is much better than the PBH bound shown.}
\label{fig-gwpbh-bound}
\end{figure}

It is seen from Fig. \ref{fig-gw-diff} that the maximal values of 2-nd order GWB amplitudes that
can be reached do not depend significantly on the width of the primordial power spectrum.
The growth of $\Omega_{GW}$ with increase of $\Sigma$, which is naturally expected
(see \cite{Bugaev:2009zh}), is partly compensated by the simultaneous
decrease of the curvature perturbation amplitude ${\cal P}_{\cal R}^0$ (really,
Fig. 2 shows that constraint on ${\cal P}_{\cal R}$ decreases with the growth of $\Sigma$).
This allows to put a constraint on the value of $\Omega_{GW}$, which is independent on the width.
Such constraint is shown in Fig. \ref{fig-gwpbh-bound}
for the whole range of GW frequencies considered, $\sim 10^{-10} \div 10^4\;$Hz.

The shaded triangle in Figs. \ref{fig-gw-diff}a, \ref{fig-gwpbh-bound} designates the pulsar
timing limit obtained in \cite{Thorsett:1996dr},
\begin{equation}
\label{ThorsettLimit}
\Omega_{GW}(f)h^2 < 4.8\times 10^{-9} \left( \frac{f}{4.4 \times 10^{-9} \; {\rm Hz} } \right)^2,
\end{equation}
for $f > 4.4 \times 10^{-9}\; {\rm Hz}$ at 90\% C.L.
It is seen that PBH mass spectra from Fig. \ref{fig-pbh-sp} ($M_h^0=0.5 M_\odot)$
are inconsistent with this limit: the same scalar perturbation spectrum overproduces GWs.
We have estimated that the limit (\ref{ThorsettLimit}) excludes the significant
amount of PBHs in the region of masses $\sim (0.03 \div 10 ) M_\odot$ due to overproduction
of 2-nd order GWs. Note that this conclusion differs from the earlier result of \cite{Saito:2008jc}
who associate the pulsar timing limit (\ref{ThorsettLimit}) with PBHs of mass $\sim 10^3 M_\odot$.
This is obviously caused by the definition $f\equiv 2\pi c k$ used by these authors, instead of
the usual one, given by Eq. (\ref{fck}). From our results it follows that the primordial
origin of intermediate mass
black holes (IMBHs), with masses $\sim (10^{2} - 10^{4}) M_\odot$, is not excluded by
pulsar timing limits.

PBHs with mass close to $\approx 10^{17}\;$g can be responsible for cosmic dark matter, and,
as shown in \cite{Bambi:2008kx, Bugaev:2008gw}, if they are clustered in the Galactic Center, they can explain the
511-keV photon line observed from its direction (such photons, in this case,
are produced by the annihilation of positrons evaporated from PBHs \cite{Okele, Frampton:2005fk}).
It is seen, in particular,
from Fig. \ref{fig-gw-diff}b that to explain this phenomenon with clustered PBHs,
$\Omega_{GW}$ must be approaching a value of $\approx 10^{-7}$ near $f \approx 1\;$Hz.
This region will be probed in future by BBO experiment.

Currently, the analysis of pulsar timing data is the only experiment which
allows to set a stronger bound on primordial spectrum than PBHs (see also \cite{Assadullahi:2009jc}).
For comparison we have also shown the limit obtained by the ground-based interferometer LIGO during
its fifth science run (S5) \cite{Nature},
\begin{equation}
\Omega_{GW} < 6.9 \times 10^{-6}. \label{LIGOS5}
\end{equation}
This limit applies to a scale-invariant GW spectrum in the frequency range $41.5-169.25$ Hz.
The target sensitivity \cite{Abbott:2006zx} of the planned Advanced LIGO
experiment, $\Omega_{GW} \sim 10^{-8} \div 10^{-9}$, is also shown in
Fig. \ref{fig-gwpbh-bound}. It is seen that in future the obtained limit for GWB will
be experimentally approached in this experiment, and also in most other regions of the frequency range considered,
by other experiments such as LISA \cite{lisa}, Big Bang Observer (BBO, see, e.g., \cite{Crowder:2005nr}),
Square-Kilometer-Array (SKA, see, e.g., \cite{Kramer:2004rwa}).

To show the uncertainty in PBH constraints, we have also plotted in Fig. \ref{fig-gwpbh-bound}
the resulting constraints obtained assuming a somewhat larger PBH formation threshold, $\delta_{th}=0.45$,
and maximal GWB corresponding to $\Omega_{PBH}=10^{-5}$ (and $\delta_{th}=1/3$).

\bigskip

\bigskip

\section{Summary and discussion}
\label{sec-concl}

We have performed calculations assuming that PBHs and 2-nd order GWs are produced at times
not very close to the beginning of the radiation era $t_i$. If production of GWs and
PBHs takes place near $t_i$ (i.e., right after the end of inflation),
the limits derived in our paper can potentially be altered. One such example is a particular
case of the running mass inflation model (proposed in \cite{Stewart:1996ey, Stewart:1997wg} and further
studied in many papers (see \cite{Bugaev:2008bi} and references therein)), which predicts a
rather strong scale dependence of the spectral index, possibly
leading to largest values of ${\cal P}_{\cal R}(k)$ just near $k_i \; (\approx k_{end})$. The analysis
performed in \cite{Bugaev:2009zh}, however, shows that the maximum values of GWB amplitude that
can be reached in this case are also very close to maximum values derived in this paper.

In summary, we have performed simultaneous calculations of PBH mass spectra and induced GW background,
obtaining the constraints on values of ${\cal P}_{\cal R}(k)$ and $\Omega_{GW}(k)$
from known limits on the PBH concentration in various cosmological scale ranges.
We have explored the dependence of these limits on the shape of the primordial
spectrum (in particular, on its width). It was shown that though constraints on the peak width
may significantly depend on the shape of the spectrum, the maximal possible values
of GWB are almost unsensitive to it. This allowed us to place quite model-independent limits on
induced GWB in the wide frequency range.
We have discussed the applicability of currently available experimental data, in particular,
pulsar timing limits, to the constraining of PBH abundance. We have shown that the primordial
origin of IMBHs is not forbidden by the pulsar timing limits.

Comparing our results with the previous results of \cite{Josan:2009qn} and \cite{Saito:2009jt},
one can see that our constraints for ${\cal P}_{\cal R}^0$ almost coincide with those of
Ref. \cite{Josan:2009qn} in cases when the width of the peak in primordial spectrum is large,
$\Sigma=3, 5$ (see Fig. \ref{fig-PR0-bound}); our constraints for $\Omega_{GW}$ are
systematically weaker than those of Ref. \cite{Saito:2009jt}. Also, in our case the dependence
of $\Omega_{GW}$-constraint on the shape of the primordial spectrum is much smaller than the
analogous dependence in Ref. \cite{Saito:2009jt}.


\end{document}